\documentclass[preprint,
preprintnumbers,amsmath,amssymb,superscriptaddress]{revtex4-1}

\usepackage{lineno}
\usepackage{fullpage,amsthm,amsmath,amsfonts,bm,times,color,bbm}
\usepackage{graphicx}
\usepackage[english]{babel}
\usepackage{mathtools}
\usepackage{float}
\usepackage{wrapfig}

\usepackage{subcaption}
\usepackage{hyperref}
\usepackage{amsmath}
\usepackage{amssymb}
\usepackage{multirow}
\usepackage{booktabs}
\usepackage{makecell}

\usepackage{soul} 

\def\deg{\hbox{$^\circ$}}

\makeatletter
\begin{document}

\title{
Dynamic Arctic weather variability and connectivity
}
\author{Jun Meng}
\affiliation{School of Science, Beijing University of Posts and Telecommunications, Beijing 100876, China}
\affiliation{Potsdam Institute for Climate Impact Research, Potsdam 14412, Germany}

\author{Jingfang Fan}
\email{jingfang@bnu.edu.cn}
\affiliation{School of Systems Science/Institute of Nonequilibrium Systems, Beijing Normal University, Beijing 100875, China}
\affiliation{Potsdam Institute for Climate Impact Research, Potsdam 14412, Germany} 

\author{Uma S Bhatt}
\affiliation{Geophysical Institute, University of Alaska Fairbanks, Fairbanks, AK 99775, USA} 
\author{J\"urgen Kurths}
\affiliation{Potsdam Institute for Climate Impact Research, Potsdam 14412, Germany}
\affiliation{Geophysical Institute, University of Alaska Fairbanks, Fairbanks, AK 99775, USA} 
\affiliation{Institute of Physics, Humboldt-University, Berlin 10099, Germany}
 
\date{\today}

\begin{abstract}
The rapidly shrinking Arctic sea ice is changing weather patterns and disrupting the balance of nature. Dynamics of Arctic weather variability (WV) plays a crucial role in weather forecasting and is closely related to extreme weather events. Yet, assessing and quantifying the WV for both local Arctic regions and its planetary impacts under anthropogenic climate change is still unknown. Here, we develop a complexity-based approach to systematically evaluate and analyze the dynamic behaviour of WV. We reveal that the WV within and around the Arctic is statistically correlated to the Arctic Oscillation at the intraseasonal time scale. We further find that the variability of the daily Arctic sea ice is increasing due to its dramatic decline under a warming climate. Unstable Arctic weather conditions can disturb regional weather patterns through atmospheric teleconnection pathways, resulting in higher risk to human activities and greater weather forecast uncertainty. A multivariate climate network analysis reveals the existence of such teleconnections and implies a positive feedback loop between the Arctic and global weather instabilities. This enhances the mechanistic understanding 
of the influence of Arctic amplification on mid-latitude severe weather. Our framework provides a fresh perspective on the linkage of complexity science, WV and the Arctic.
\end{abstract} 
\maketitle

\begin{figure}
    \centering
    \includegraphics[width=1.0\linewidth]{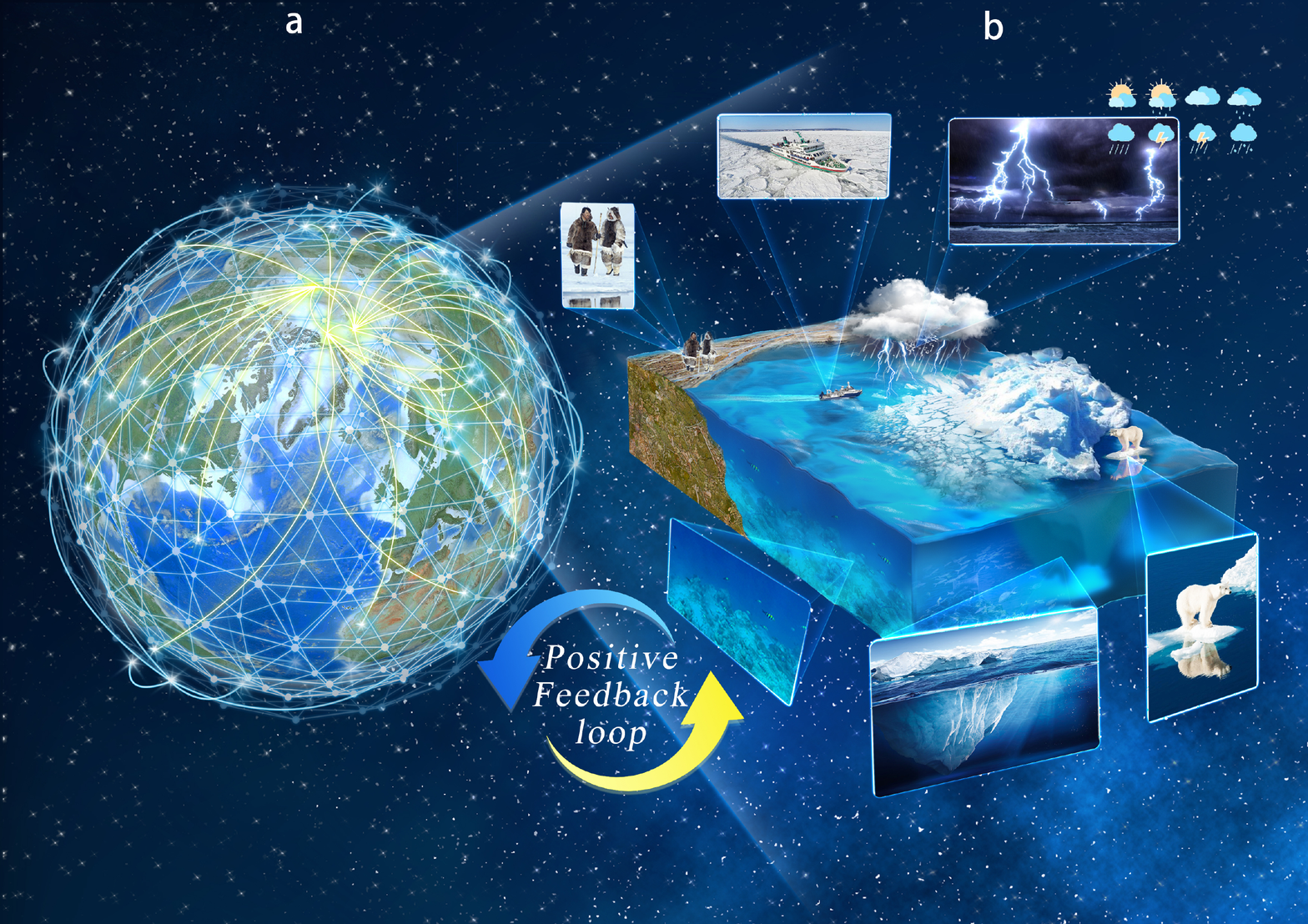}
    \caption{\textbf{The arctic system as a crucial component of the Earth climate system}. \textbf{a}, Schematic view of a climate network. Links indicate interactions between different regional climate systems in the globe. Golden links represent teleconnections between the Arctic and regions outside. \textbf{b}, Illustration of the complex Arctic system. It contains the cryosphere, biosphere, hydrosphere, and atmosphere as well as the interactions among them. A change in one component often triggers changes and feedbacks in numerous interconnected processes (e.g., Arctic sea ice decline). The circular arrow suggests a positive feedback of the WV between the Arctic and the rest of the climate system.} 
    \label{intro}
\end{figure}

Arctic sea ice is declining and thinning at an
accelerating rate due to anthropogenic climate change ~\cite{rothrock_thinning_1999,comiso_accelerated_2008}. The warming trend is more prominent in the Arctic and is double of the global average or even greater regionally~\cite{rantanen2022arctic}, a phenomenon known as Arctic amplification (AA)~\cite{miller2010arctic,serreze_processes_2011,previdi2021arctic}. The Arctic sea ice conditions can affect the Arctic ecosystem, wildlife, hunting and shipping, exploration of nature resources and more ~\cite{orbaek2007arctic,peeken_arctic_2018,schneider2021consequences}. As one crucial component of the complex Earth system~\cite{steffen_emergence_2020,assessment2005arctic}, changes in Arctic sea ice are found to have statistical and dynamical connections with regional as well as remote climatic impacts~\cite{sevellec_arctic_2017, screen_consistency_2018, chemke_effect_2019, blackport_minimal_2019} (as shown in Fig. \ref{intro})
through both large-scale atmospheric and oceanic circulations~\cite{rahmstorf_thermohaline_2003,budikova_role_2009,kushnir_atmospheric_2002,alexander_atmospheric_2004,honda_influence_2009}. The rapid shrinking of the ice cover has attracted much attention about the Arctic sea ice teleconnections and predictions from seasonal-to-decadal time scales in recent years~\cite{francis_evidence_2012,cohen_recent_2014,guemas_review_2016,olonscheck_arctic_2019}. However, the understanding about its variability on weather time scales is still in its infancy~\cite{smith_sea_2016,mohammadi-aragh_predictability_2018}, although it is crucial for weather forecasting, the safety of commercial and subsistence maritime activities, the survival of polar mammals and the benefit of polar economics. The impact of day-to-day Arctic sea ice variations has been underestimated in most of the climate models~\cite{dammann_impact_2013}. To fill this gap, here we adopt complexity-based approaches and the \textit{climate network} framework to investigate the daily WV of the Arctic sea ice and its connections to climate phenomena on different spatio-temporal scales, including the Arctic Oscillation (AO), climate change and local weather conditions even in regions faraway.

Complexity science employs the mathematical representation of network science and provides a powerful tool to study the structure, dynamics and function of complex systems~\cite{newman_networks_2010}. The climate system is a typical complex adaptive system due to its nonlinear interactions and feedback loops between and within different layers and components. In recent years, network science has been implemented to the climate system to construct the climate network (CN)~\cite{tsonis_architecture_2004}. The CN is a novel tool to unveil and predict various important climate mechanisms and phenomena~\cite{fan_statistical_2021}, including forecasting of the El Ni\~{n}o Southern Oscillation~\cite{ludescher2013improved,meng2017percolation} and Indian summer monsoon rainfall~\cite{stolbova2016tipping,fan2022network}, the global pattern of extreme-rainfall ~\cite{boers2019complex}, the changes of global-scale tropical atmospheric circulation under global warming~\cite{fan2018climate}, teleconnections among tipping elements in the Earth system \cite{liu_teleconnections_2023}, the Indian Ocean Dipole \cite{lu2022early} and so on.

The AO is one of the major modes of atmospheric circulation over the mid-to-high latitudes of the Northern Hemisphere (NH)~\cite{thompson_arctic_1998}, which influences climate patterns in Eurasia, North America, Eastern Canada, North Africa, and the Middle East, especially during boreal winter~\cite{deser_teleconnectivity_2000, rigor_response_2002,he_impact_2017}. The AO index is defined as the leading empirical orthogonal function of NH sea level pressure anomalies from latitudes $20^\circ$ N to $90^\circ$ N and is characterized by the back-and-forth shifting of atmospheric pressure between the Arctic and the mid-latitudes. During the positive AO phases, the surface pressure is lower-than-average in the Arctic region and the jet stream shifts northward accompanied by a poleward shift of the storm track~\cite{simmonds_arctic_2008}. Correspondingly, we find that both the sea ice and air temperature in mid-to-high latitudes of the NH changes more rapidly (i.e., with blueshifted frequency spectrum) paired with more stable weather conditions (i.e., redshifted) in regions further south during the \textit{AO positive phases}, in contrast to the \textit{AO negative phases} when pressure north of the Arctic Circle is higher than normal. To quantify the blue/red-shift effect and its geographic distribution indicating increased/reduced WV, here we introduce two novel mathematical techniques: the \textit{advanced autocorrelation function method}, i.e., $W_{ACF}$ and the \textit{advanced power spectrum method}, i.e., $W_{PS}$ (see Methods). This way enables us to find that the day-to-day variability of ice cover for a large area of the Arctic is increasing due to the dramatic melting of the sea ice~\cite{kwok2018arctic}, which indicates more enhanced risks for severe weather under climate change~\cite{zhang_climatology_2004,yin_consistent_2005,valkonen_arctic_2021,parker2022influence}. This may also increase the probability of unstable weather conditions globally through atmospheric teleconnections between the Arctic and the global climate systems (see links shown in Fig. \ref{intro}). Finally, we statistically verify the existence of such teleconnections between the Arctic sea ice and weather conditions in remote global regions via a multivariate climate network framework. Such teleconnections can result in a positive feedback loop of WV between the Arctic and the rest (see Fig. \ref{intro}) and contribute to understanding the mechanisms of linkage between the AA and mid-latitude weather~\cite{cohen2020divergent}. The presented results and methodology not only facilitate a quantitative risk assessment of extreme weather events (see Fig. S1), but also reveal the existence of interaction or synchronization paths among regional and global climate components.

\section*{Results}
\subsection*{Linkage of the weather variability and the AO}
The WV refers to the irregularity/predictability of the climate data at weather time scales (i.e., hours - days). There are various ways to evaluate the data variability/irregularity, such as the entropy ~\cite{pincus1995approximate,richman2000physiological,costa2005multiscale}, the detrended fluctuation analysis~\cite{peng1995quantification,livina2007modified}, the correlation dimension~\cite{procacia1983measuring}, the lyapunov exponents analysis~\cite{wolf1985determining}, etc. However, most of them would be problematic, biased or invalid when dealing with short and noisy data, such as the weather data. The standard deviation (SD) is an effective technique to quantify the dispersion of data, but not a good measure for irregularity, e.g., the SD of a randomly shuffled data is the same as the original. Besides, the auto-correlation function describes how fast the self-similarity of a variable decays with time~\cite{box_time_2015} and the power spectral analysis~\cite{stoica2005spectral} allows us to discover periodicity in the data. Yet, a systematic evaluation of the auto-correlation and the power spectrum as well as their dynamic evolution for non-stationary climate data are still lacking. 

Therefore, here we introduce two mathematical functions: $W_{ACF}$ and $W_{PS}$ (see Methods for details) to quantify the WV in and around the Arctic in a given month, as well as its dynamic behavior during the period from Jan. 1980 to Dec. 2019. For a given time series, the physical meanings of these metrics are:  higher values of the $W_{ACF}$ stands for weaker short-term memory; while higher values of the $W_{PS}$ indicates faster changes.
In particular, to better understand their physical meanings, we construct various nonlinear time series (as shown in Fig. \ref{example}a) via the following dynamical equations,
\begin{align}
x_t = \cos{(2\pi t/20)},        \label{eq1}\\
y_t = \cos{(2\pi t/10)},         \label{eq2}\\
z_{t}^{x} = 0.2x_t + 0.8u_t,     \label{eq3}\\
z_{t}^{y} = 0.2y_t + 0.8u_t,     \label{eq4}
\end{align}
where $t\in [0,1000]$, $u_t$ is the nonlinear logistic function as: $u_{t+1} = \mu u_{t} (1-u_{t})$. Here we set the parameter $\mu = 3.8$ and $u_0 = 0.01$, i.e., it generates a chaotic behavior~\cite{pomeau_intermittent_1980}. Mathematically, Eqs.~(\ref{eq1}) and (\ref{eq2}) are two periodic functions but with different periods $20$ and $10$, respectively; while, Eqs.~(\ref{eq3}) and (\ref{eq4}) consist of a periodic term and a chaotic term (Fig. \ref{example}a). Therefore, strictly speaking, the value of $W_{ACF}$ for $z_{t}^{x}$ ($z_{t}^{y}$) is higher than $x_t$ ($y_t$), i.e., weaker short-term memory, due to the chaotic term $u_t$; the value of $W_{PS}$ for $y_{t}$ ($z_{t}^{y}$) is higher than $x_t$ ($y_t$), i.e., faster changes, due to the periodic term with different periods. 
One should note that a segment of unstable data is usually  changing faster, with both high $W_{ACF}$ and $W_{PS}$, e.g., Eqs. (\ref{eq3}) and (\ref{eq4}). While a segment of quickly changing data is not necessarily irregular, such as high frequency periodic data, with high $W_{PS}$ but low $W_{ACF}$, as Eq. (\ref{eq2}). We extract 31 (i.e., the maximal length of one month in the climate data) consecutive data points from each of the samples and perform $W_{ACF}$ and $W_{PS}$ analysis on the extracted subsets. All results presented in Fig. \ref{example}b and c, are consistent with our theory, which indicates that our two functions can be used as effective tools to describe the variability (both \textit{disorder} and \textit{frequency}) for given time series. 

Next, we apply $W_{ACF}$ and $W_{PS}$ to quantify the Arctic sea ice WV based on the sea ice cover dataset (daily, 1979-2019, see DATA for details). Our results are shown in Figs. \ref{example}d-h.
A positive value of $r$ denoted by blue in Figs. \ref{example}d or e, indicates positive correlation between the annual mean of $W_{ACF}$ or $W_{PS}$ with the AO index. We observe that both $W_{ACF}$ and $W_{PS}$ tend to be higher, i.e., indicating faster and more irregular day-to-day changes of ice cover, during the \textit{AO positive phases} than \textit{AO negative phases}, in some parts of the Arctic region, such as, the Canadian Archipelago, Beaufort Sea, and the Central Arctic. To illustrate the effect of the AO on $W_{PS}$, we show that the power spectrum of Arctic sea ice during the AO positive phase, e.g., Jan. 1989, is significantly blueshifted comparing to that during the AO negative phase, e.g., Jan. 2010 (see Fig.~\ref{example}f). To illustrate the effect of the AO on $W_{ACF}$, we show that the timeseries of AO index and $W_{ACF}$ are significantly synchronized during the period 1980-2019 (as shown in Fig. \ref{example}g and h). Moreover, we uncover that the climatic effects of the AO are more prominent in winter-spring than in summer-autumn (see Figs. S2 and S3).

The underlying physical mechanism is related to the typical atmospheric character of the AO, as well as the close interactions between the Arctic sea ice and the surface atmosphere. During the positive phases of AO, the jet stream shifts northward and the storm tracks are located farther north than during the AO negative phases~\cite{thompson_regional_2001}, see Fig. S4. This results in more unstable regional weather in mid-to-high latitudes of the NH, and yields higher $W_{ACF}$ and $W_{PS}$ of the air temperature data, see Fig.~\ref{corrAO} and Figs. S5-S8.
In contrast, the $W_{ACF}$ and $W_{PS}$ of the air temperature in the mid-latitudes of the NH increase with more outbreaks of significant weather events (e.g., cold events, frozen precipitation and blocking days)~\cite{thompson_regional_2001} as the zonal wind weakens during the negative AO phases, see Fig.~\ref{corrAO} and Figs. S5-S8. In particular, as shown in Figs. S5-S8, there are even significant connections between the AO and the WV in some regions of the Southern Hemisphere. 

The $W_{ACF}$ and $W_{PS}$ analysis provides an additional way to describe the quantitative response of both the Arctic sea ice and the atmosphere to the AO, thus could be used to assess the risk of extreme events in mid-to-high latitudes of the NH.

\subsection*{Increased irregularity of Arctic sea ice cover}

In the following, our results shown in Fig. \ref{trends} indicate that the sea ice cover in a large area of the Arctic, including the East Siberian, the Beaufort Sea and the Central Arctic, where the ice thickness decrease is dramatic (as shown in Fig. S9), has changed more rapidly and irregularly over the past 40 years (1980-2019). That is since both values of the $W_{ACF}$ and $W_{PS}$ are significantly increasing. The observed enhancing trend of WV may be attributed to the following two reasons: One is related to the development of remote sensing and data analyzing technology, resulting in better data resolution and accuracy over the data record; the other reason is the rapid decline of multi-year ice cover, due to the dramatic increase of air temperature~\cite{comiso_large_2012}. 
The multi-year sea ice has been defined as the ice that survives at least one summer melt and represents the thick sea ice cover, while the first-year ice refers to the ice that has no more than one-year's growth. As more  of perennial ice cover is replaced by younger and thinner ice cover, the regional ice cover becomes more fragile and vulnerable to fluctuations of air temperature or some other forces~\cite{kwok2018arctic}. Therefore, local interactions between the sea ice and atmosphere would be enhanced and the weather in the Arctic and remote global regions may affect each other more easily through potential tele-connected pathways (e.g., Fig. \ref{PATH}), which may increase the WV associated with the short-term weather predictability.

In addition, we observe relatively more areas with a significant trend of enhanced instability in the melt season under global warming (see Fig. \ref{trends}a). This is because during the melt season (Apr.-Aug.), the sea ice declines and fluctuates more dramatically than in other seasons when the monthly average ice cover extent (the area of ocean with at least $15\%$ sea ice, marked by the blue curve in Fig. \ref{trends}a) reaches its maximum/minimum. An intensification of the summer Arctic storm activity is also likely to happen as the land-sea thermal contrast increases under global warming ~\cite{day2018growing,kenigson_arctic_2021,peng_role_2021}, which can increase the WV both in the ocean and atmosphere.

\subsection*{Arctic-global teleconnection patterns}
Next, we propose the \textit{multivariate climate network} approach to statistically reveal the potential teleconnection patterns between the Arctic sea ice (Fig. S10a) and the global air temperature field (Fig. S10b), see more details in the Methods. Different from  the classical climate network approach with only one climate variable, see Ref. \cite{donges_unified_2015,dijkstra2019networks,fan_statistical_2021} and references therein, we construct climate networks where each link connects one node located in the Arctic (Fig. S10a) and the other in the globe (Fig. S10b). In particular, the link weight quantifies the similarity of temporal evolution between two different climate variables, i.e., the Arctic sea ice and the global air temperature. By comparing to a Null-model (see Methods), we observe the dynamic behavior of network connectivity (as shown in Fig. S11a), which is defined as the ratio of significant links for each month's network. The statistical significance for each link is defined by comparing to the null-model, see details in Method section. A value of above $5\%$ connectivity indicates statistically significant synchronization of weather between the Arctic and areas outside, such as Feb. 2010 (see Fig. S11b and c), when the AO is in a strong negative phase and the cold polar air plunged into lower latitudes of the NH and result in extreme weather conditions in a large area of the globe~\cite{cohen2010winter,liu2012impact}.
We identify the significant Arctic-global teleconnection patterns by using climate network node degree fields, which are defined as the number of significant links that connect to the Arctic for each global node, for two  specific periods,  Feb. 2010 (AO negative phase) and Mar. 2019 (AO positive phase) in Fig. \ref{PATH}a and c, respectively.

Moreover, two typical links presented in Fig.~\ref{PATH} indicate strong synchronizations between the daily sea ice cover for one Arctic node and the air temperature for another remote global node (their time series are shown in Figs. S12 and S13). 
As shown in Fig. S12b, changes in the sea ice for node $i$ ($77.5\deg$ N, $160\deg$ E) in the Arctic are two days ahead of the air temperature variations for node $j$ ($30\deg$ N, $105\deg$ E) in the Sichuan Province of Southwest China, i.e., evolution of the Arctic sea ice could affect the anomalies of air temperature in Southwest China. To better understand how sea ice affects air temperature variability faraway, we identify the most probable teleconnection propagation path through the \textit{shortest path} method (see Methods for more details). We show a potential propagation path for this teleconnection (marked by yellow in Fig. \ref{PATH}b) and find that it seems to be roughly a straight line from the Arctic to Southwest China through Eastern Russia and Mongolia. The path length is close to $6400$ km. From a meteorological perspective, this path can be well explained by the main large-scale atmospheric circulation. A negative phase of the AO leads to a stronger Siberian High and extends farther southeastward. This results in repeated cold air outbreaks into South China \cite{he_impact_2017}. Our analysis is highly consistent with the wind climatology, see the background information of Fig. \ref{PATH}b.

In addition, its feedback is also considered. However, we observe a relatively weaker connection in the opposite direction, i.e., from Southwest China to the Arctic. We find that changes in the air temperature at the same location in Southwest China influence that of the sea ice for the same Arctic node $11$ days later, as shown in Fig. S12c. Correspondingly, we identify its potential propagation path (marked by orange in Fig. \ref{PATH}b) and find it corresponds to negative wind anomalies from Southwest China to the Arctic. These two tele-connected paths form an interaction loop that suggests a large-scale atmospheric feedback of WV between the Arctic and Southwest China.

In a contrast, during a positive phase of the AO, we show another teleconnection and its path in Fig. \ref{PATH}c and d, which indicates that the fluctuations of air temperature in California can affect the Arctic sea ice through the upper atmospheric circulations. Meanwhile, changes in the Arctic sea ice can also influence the temperature fluctuations in California along upper wind routes in an opposite direction, however, at a weaker strength (see more details in Fig. S13c). This is because during the positive phase of the AO, low pressure dominates the Arctic regions, leading to a northward and intensified jet stream that blocks the outbreaks of frigid polar air into lower latitudes and reduces storm activity in California~\cite{stroeve_arctics_2012}. The uncovered teleconnection loop between the Arctic and California suggests that Arctic sea ice decline may drive more California droughts and wildfires ~\cite{cvijanovic_future_2017}.

The synchronization of day-to-day weather between the Arctic and other regions can favor positive feedbacks of WV, where increasing WV/instability of the Arctic sea ice may cause a higher risk of extreme weather conditions in remote global regions. Meanwhile, impacts from global regions may also induce unstable weather conditions in the Arctic.

\section*{Discussion}
In summary, we have introduced the mathematical $W_{ACF}$ and $W_{PS}$ functions to quantify the short-term dynamic WV relating to the irregularity and frequency of the day-to-day changes of climate data. By adapting $W_{ACF}$ and $W_{PS}$, we are able to identify significant effects of the AO on day-to-day changes of the Arctic sea ice as well as the WV in mid-to-high latitudes of the NH. We attribute the physical mechanism to the shifts of north-to-south location of jet stream and storm-steering associated with different phases of the AO. Furthermore, we found that during the past 40 years, the Arctic sea ice variability on weather time scales is substantially increasing due to the melting of the thick perennial sea ice. Finally, in order to analyze the dynamic Arctic weather connectivity, we have constructed multivariable climate networks, i.e., between the Arctic sea ice and the global air temperature field. By applying the shortest path method, we are able to identify teleconnections paths as well as positive feedback loops of WV. We also proposed a possible physical mechanism underlying these paths. The reduction of Arctic sea ice stability may increase the risk of unstable weather conditions and lead to reduced skill of weather forecasts~\cite{petropoulos_forecasting_2022} globally through the Arctic-global teleconnected feedback loops. Our new findings can help to understand the physical mechanisms linking the AA and the global climate, and implies prominent global impacts of the Arctic WV on human and natural systems under climate change~\cite{cohen2020divergent,previdi2021arctic}.

As the Arctic is considered to be a barometer of global climatic change, in particular, Arctic sea ice loss is approaching a tipping point and is extremely crucial for the whole Earth’s climate \cite{lenton_climate_2019}. Besides the immediate utility of being able to quantitatively analyze the dynamics of WV for local Arctic regions and its global impacts, our framework would be also applied to study and reveal the short-term synchronizations of connectivity among remote global regions, sea ice forecasting, as well as systemic risk induced by the interdependency among other complex subsystems and cascading of adverse consequences, which is particularly important for a systemic risk-informed global governance.

\clearpage

\clearpage

\begin{figure}
\begin{centering}
\includegraphics[width=0.85\textwidth]{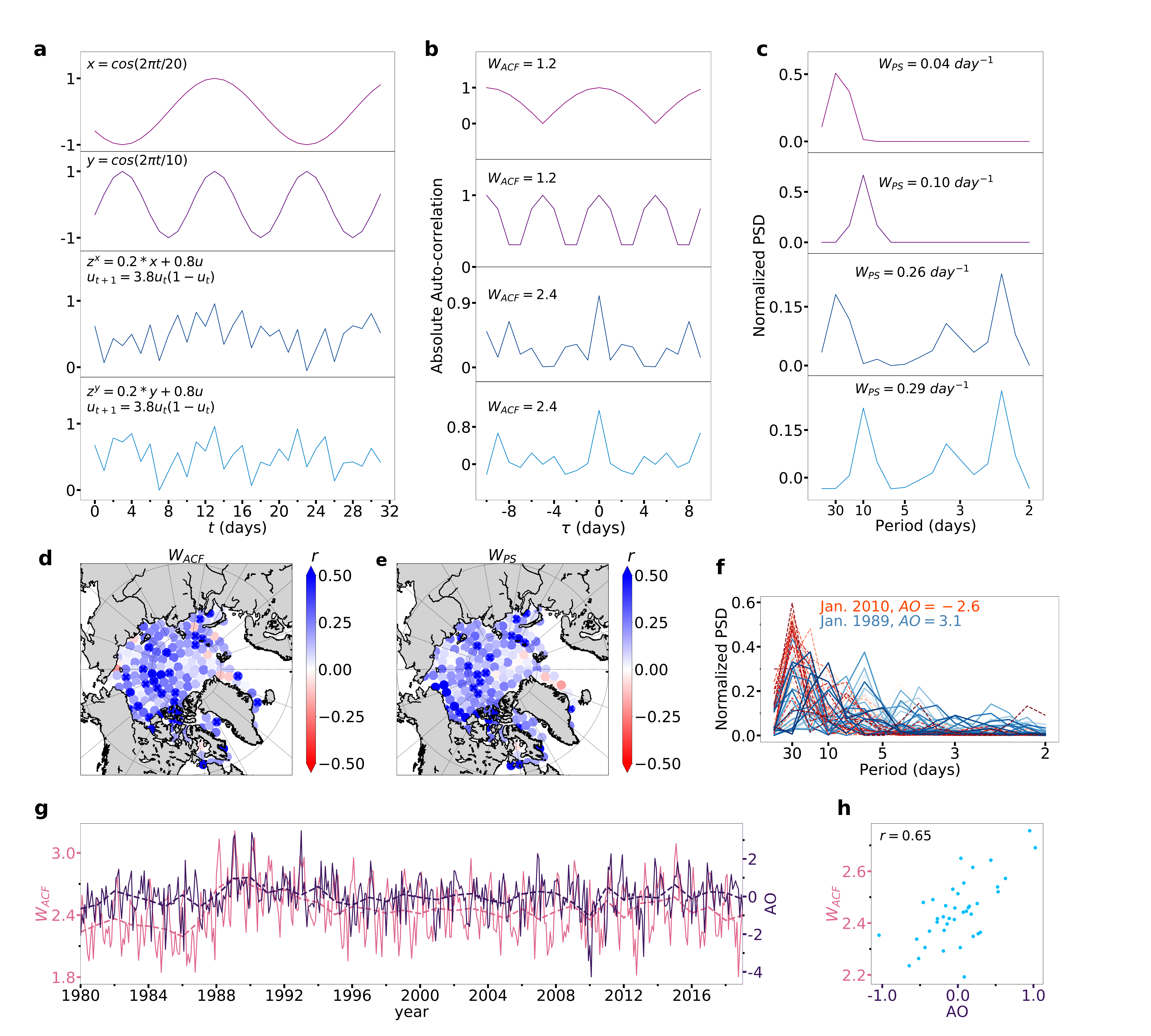}
\caption{\label{example}\textbf{Blueshift effect of the Arctic Oscillation on the Arctic weather variability}. \textbf{a}, Sample nonlinear time series generated based on Eqs. (\ref{eq1}-\ref{eq4}). \textbf{b}, The auto-correlation functions and values $W_{ACF}$ of each sample time series shown in \textbf{a}. \textbf{c}, The power spectrum density and values $W_{PS}$ of each sample time series shown in \textbf{a}. \textbf{d}, The correlations between the annual mean of the AO index and the $W_{ACF}$ for the Arctic sea ice. 
The ``\textbf{x}" marks represent the nodes with correlations significant at the 95\% confidence level (Student’s t test). \textbf{e}, The same as \textbf{d} for $W_{PS}$. \textbf{f}, The power spectrum of the sea ice for all nodes marked by symbol ``\textbf{x}" in \textbf{e} in Jan. 1989 with a positive AO phase comparing to that in Jan. 2010 with a negative AO phase. \textbf{g}, The AO index (pink solid line for monthly and pink dashed line for annual) versus the $W_{ACF}$ index (dark blue solid line for monthly and dark blue dashed for annual) averaged over all nodes marked by symbol ``\textbf{x}"  in \textbf{d}. \textbf{h}, The scatter plots of annual indexes (dashed lines in \textbf{g}) of the AO versus $W_{ACF}$, the $r$ value between these two indexes is $0.65$, with a $p$ value of $5.5\times 10^{-6}$.}
\par\end{centering}
\end{figure}

\clearpage

\begin{figure}
\begin{centering}
\includegraphics[width=0.9\textwidth]{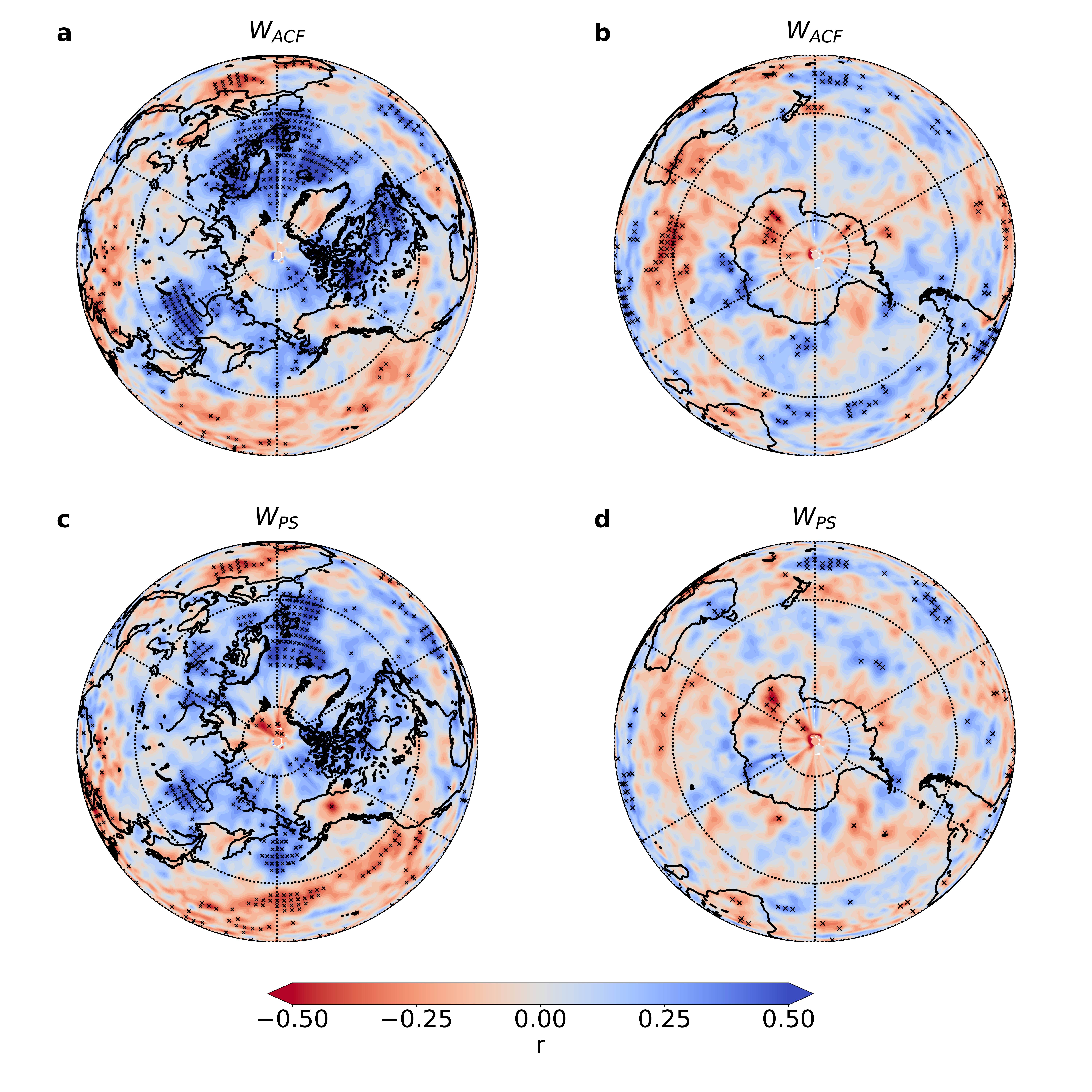}
\caption{\label{corrAO}\textbf{
The relationships between the AO and weather variability.}  \textbf{a},\textbf{b}, The correlation maps between the annual mean of the AO index and $W_{ACF}$ of the air temperature at $850hPa$ pressure level during the period of 1980--2019. \textbf{c},\textbf{d}, The same as \textbf{a} and \textbf{b}, but for $W_{PS}$. The symbol ``\textbf{x}" in each panel
represents the region with correlation significant at the 95\% confidence level (Student’s t-test).}
\par\end{centering}
\end{figure}

\clearpage

\begin{figure}
\begin{centering}
\includegraphics[width=1.0\textwidth]{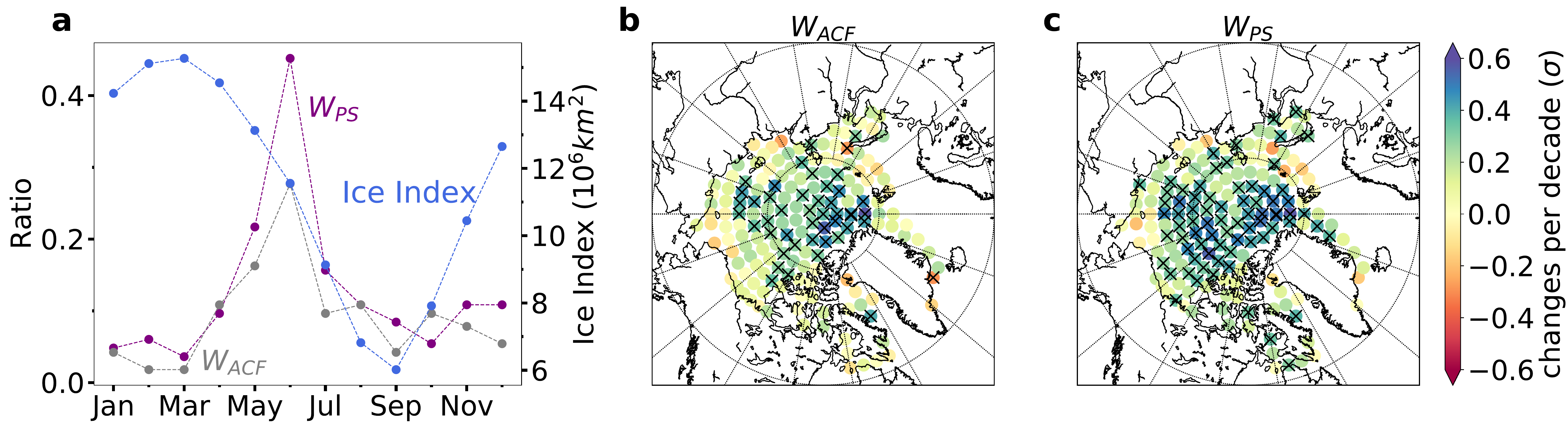}
\caption{\label{trends}\textbf{The dynamic weather variability of the Arctic daily sea ice cover during June.}  \textbf{a}, The ratio of nodes that has statistically significant increasing trend for the $W_{ACF}$ (gray) and $W_{PS}$ (purple); the Sea Ice Index, i.e. the area with at least $15\%$ ice cover (blue) for the same months during 1980--2019. \textbf{b}, Changes per decade as multiple of one standard deviation ($\sigma$), for each Arctic node's $W_{ACF}$ during June. \textbf{c}, the same as \textbf{b} for $W_{PS}$. The symbol ``\textbf{x}" in panels \textbf{b} and \textbf{c} represents the region with trend significant at the 95\% confidence level (Student’s t-test).}   
 \label{strength}
\par\end{centering}
\end{figure}

\clearpage

\begin{figure}
\begin{centering}
\includegraphics[width=0.9\textwidth]{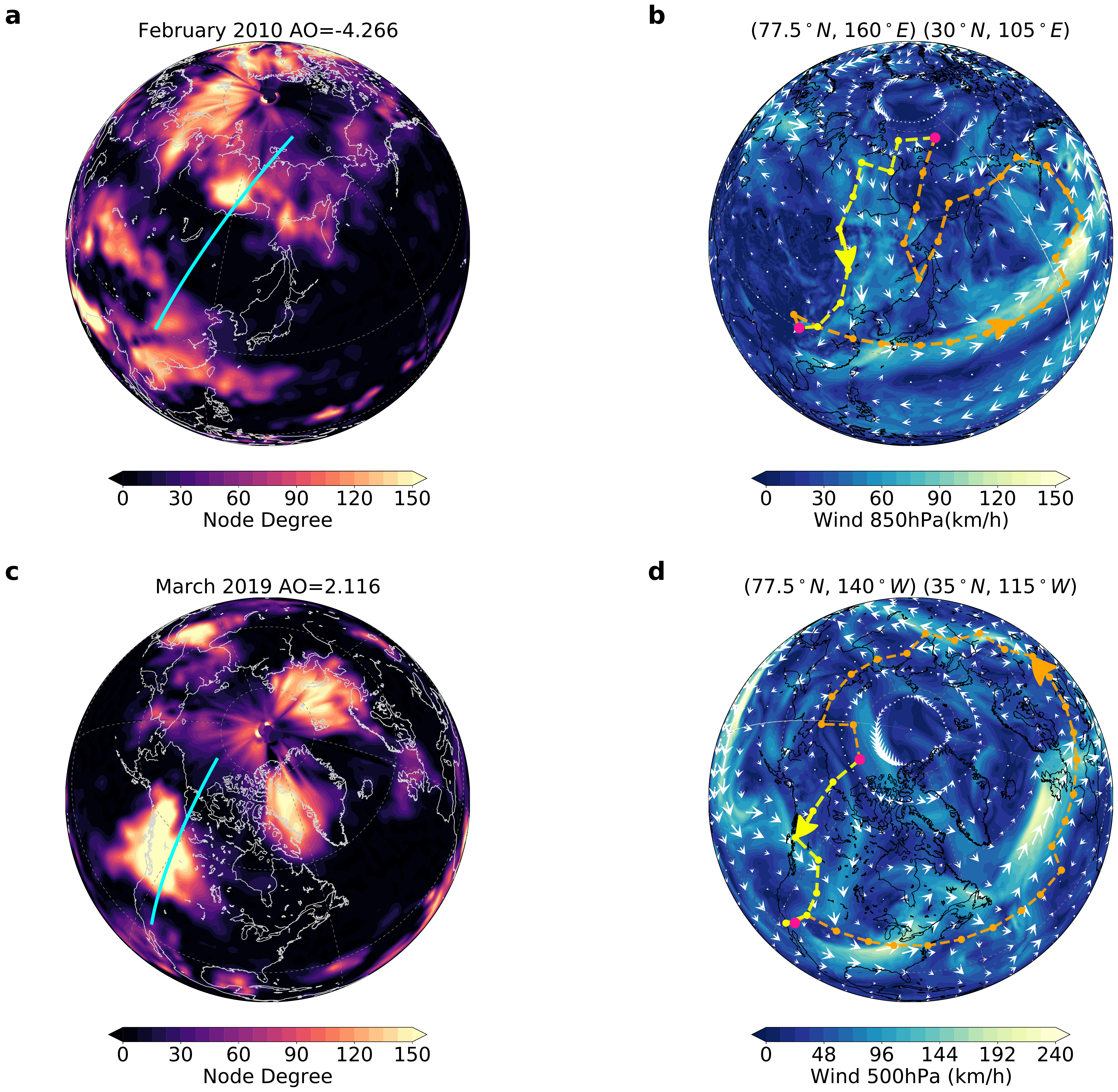}
\caption{\label{PATH} \textbf{Diagram of climate network teleconnection paths.} \textbf{a}, Heatmap of the node degree defined as the number of significant links for each node (see Methods) in the climate network of Feb. 2010. The blue line indicates the teleconnection  between one Arctic node and one node located in Sichuan province of China. \textbf{b}, The propagation pathway of the teleconnection marked by blue in \textbf{a}. \textbf{c}, the same as \textbf{a} for Mar. 2019. The blue line indicates the teleconnection link between one Arctic node and one node in California of United States. \textbf{d}, The propagation pathway of the teleconnection marked by blue in \textbf{c}. The colors and white arrows depict the magnitudes and directions
of the 850 (500)-hPa winds in \textbf{b} (\textbf{d}).
}
\par\end{centering}
\end{figure}

\clearpage

\section*{Data and Methods}
\subsection*{Data}
The data used in the current work is the $0$ hr (UTC) daily sea ice cover and the air temperature at $850 hPa$ pressure level from the ERA5~\cite{era5} (\url{https://apps.ecmwf.int/datasets/}) reanalysis, with a spatial (zonal and meridional) resolution of $2.5^\circ \times 2.5^\circ$. The searching principle for $850 hPa$ pressure level is, since it is just above the boundary layer to avoid direct interactions between the sea ice and surface atmosphere~\cite{olonscheck_arctic_2019}. We select $8040$ grids from the dataset of air temperature which approximately equally cover the globe (see Fig. S10b). There are $377$ grids located in the ocean of the Arctic region that with non-zero sea ice cover at least for one day (see Fig. S10a). Then, for each calendar year $y$ and for each node, we calculate the anomalous value for each calendar day $t$ by using the original value minus the climatological average, then divided by the climatological standard deviation. The calculations of the climatological average and standard deviation are based on data from the year of 1979 to 2019. For simplicity, leap days are excluded.

The AO index was downloaded from: \url{https://www.cpc.ncep.noaa.gov/products/precip/CWlink/dailyaoindex/monthly.ao.index.b50.current.ascii}. [Accessed in Sep. 2021].

The Arctic Sea Ice Extent was downloaded from :
\url{https://nsidc.org/data/g02135/versions/3}. [Accessed in Jan. 2021].


\subsection*{Assessing Weather Variability Functions}
\subsubsection*{Advanced autocorrelation function method}

The autocorrelation function (ACF) is widely used to measure the memory of a time series and reveals how the correlation between any two values of the signal changes as their time-lag \cite{box_time_2015}.  Generally, for a given time series, $x_{t}$, the ACF is defined as,
\begin{equation}\label{EQACF}
C(\mathrm{\tau})=\frac{\operatorname{Cov}\left(\mathrm{x}_{\mathrm{t}}, \mathrm{x}_{\mathrm{t}+\mathrm{\tau}}\right)}{\sqrt{\operatorname{Var}\left(\mathrm{x}_{\mathrm{t}}\right) \operatorname{Var}\left(\mathrm{x}_{\mathrm{t}+\mathrm{\tau}}\right)}},   
\end{equation}
where $\operatorname{Cov}(\bold{X}, \bold{Y})=\mathrm{E}[(\bold{X}-\mathrm{E}[\bold{X}])(\bold{Y}-\mathrm{E}[\bold{Y}])]$ and $\operatorname{Var}(\bold{X}) = \mathrm{E}[\bold{X^2}] - \mathrm{E}[\bold{X}]^2$.  If the $x_t$ are completely uncorrelated, for example, a white noise process, $C(\mathrm{\tau})$ is zero at all lags except a value of unity at lag zero ($\tau=0$). A correlated process on the other hand, has non-zero values at lags other than zero to indicate a correlation between different lagged observations.
In particular, short-range memory of the $x_t$ are described by $C(\mathrm{\tau})$ declining exponentially
\begin{equation}\label{ACF_SHORT}
 C(\tau) \sim \exp \left(-\tau / \tau^{*}\right),   
\end{equation}
with a characteristic time scale, $\tau^{*}$. For long-range memory, $C(\tau)$  declines as a power-law
\begin{equation} \label{ACF_LONG}
C(\tau) \propto \tau ^{-\gamma}, 
\end{equation}
with an exponent $0<\gamma<1$. However, a direct calculation of $C(\tau)$, $\tau^{*}$ and $\gamma$ is usually not appropriate due to noise superimposed on the collected data $x_t$ and due to underlying trends of unknown origin \cite{kantelhardt_detecting_2001}.  In order to overcome the problems described above, here, we develop an advanced autocorrelation function method to quantify the memory (both short and long range)  strength $W_{ACF}$ of a time series as,
\begin{equation} \label{W_ACF}
{W_{ACF}} = \frac{\text{max}\left(|C(\tau)|\right)-\text{mean}\left(|C(\tau)|\right)}{\sqrt{\operatorname{Var}\left(|C(\tau)|\right)}} \equiv \frac{1 -\text{mean}\left(|C(\tau)|\right)}{\sqrt{\operatorname{Var}\left(|C(\tau)|\right)}},
\end{equation} 
where `max' and `mean' are the maximum and mean values of the absolute ACF, i.e., $|C(\tau)|$, respectively. $\tau\in[-\tau_{max}, \tau_{max}]$ is the time lag. In the present work, we take $\tau_{max}=10$ days, since we are considering the day-to-day changes of data at the time scale of weather forecasting, i.e., within two weeks. Equation (\ref{W_ACF}) describes the fluctuations of the ACF and its values reveal the strength of memory, i.e., higher (smaller) $W_{ACF}$ indicates a weaker (stronger) correlation and results in a low (strong) memory.  For example,  white noise  has a maximum value $W_{ACF} =  (2\tau_{max} + 1) \sqrt{\frac{2\tau_{max}}{2\tau_{max}+1}}$. Other examples are described in Fig.  \ref{example}. Another big advancement of our method is eliminating the problematic nonstationarities.

\subsubsection*{Advanced power spectrum method}
The advanced autocorrelation function $W_{ACF}$ quantify well the strength of memory for an arbitrary time series, but does not reveal any information about the frequency content. For example, Eqs. (\ref{eq1}) and (\ref{eq2}) are two functions with different periods. Their $W_{ACF}$ values are almost the same, as shown in Fig. \ref{example}. To fill this gap, we further develop an advanced power spectrum (PS) method. Based on the Welch's method \cite{welch_use_1967} we define the advanced power spectral density $W_{PS}$ as, 
\begin{equation} \label{W_PS}
W_{PS}=\int_{f}P(f) \times f df,
\end{equation} 
where $P(f)$ is the normalized spectral density and $f$ stands for the corresponding frequency, which can be obtained by Fourier transform. $W_{PS}$ is indeed the weighted mean of $f$, thus has the same unit as frequency.  Notably, a relatively higher value of the $W_{PS}$ indicates a larger ratio of the high frequency components (i.e., blueshift), see examples shown in Fig. \ref{example}.

\subsection*{Climate Networks}
\subsubsection*{Nodes}
Different from the classical climate network with only one node classification, see Ref. \cite{dijkstra2019networks,fan_statistical_2021} and references therein, here, we define two types of nodes:  globe nodes $i$  with air temperature variable $T_{i} (t)$; Arctic nodes $j$  with Arctic sea ice cover variable $I_{j} (t)$. We thus have $8040$ globe nodes (as shown in Fig. S10b) and $377$ Arctic nodes (as shown in Fig. S10a).
\subsubsection*{Links}
We construct a sequence of multivariate climate networks. 
For obtaining the strength of the links between each pair of nodes $i$ and $j$, we compute, for each month $m$, the time-delayed, cross-correlation function
\begin{equation} \label{CNEQ1}
C_{i, j}^{m}(\tau)=\frac{\left\langle T_{i}^{m}(t) I_{j}^m(t-\tau)\right\rangle-\left\langle T_{i}^{m}(t)\right\rangle\left\langle  I_{j}^m(t-\tau)\right\rangle}{\sqrt{\operatorname{Var}(T_i^m(t)) \operatorname{Var}(I_{j}^m(t-\tau))}},
\end{equation}
 and 
\begin{equation}\label{CNEQ2}
C_{i, j}^{m}(-\tau)=\frac{\left\langle T_{i}^{m}(t-\tau) I_{j}^m(t)\right\rangle-\left\langle T_{i}^{m}(t-\tau)\right\rangle\left\langle  I_{j}^m(t)\right\rangle}{\sqrt{\operatorname{Var}(T_i^m(t- \tau)) \operatorname{Var}(I_{j}^m(t))}},
\end{equation}
where the bracket $\left\langle \right\rangle$  denotes
an average over consecutive days during a given month $m$, and $\tau\in[0,\tau_{max}]$ is the time lag.  Since we mainly focus on the dynamic Arctic WV, here we chose the maximal time lag $\tau_{max} = 20$ days for Eqs. (\ref{CNEQ1}) and (\ref{CNEQ2}).

We identify the time lag $\theta$ at which the absolute value of the cross-correlation function $|C_{i,j}^m(\tau)|$ reaches its maximum. The \textit{weight} of link $(i,j)^m$ is defined as the corresponding value of the cross-correlation function, i.e. $C_{i,j}^m=C_{i,j}^m(\tau=\theta)$. Therefore, the weight of each link could be either positive or negative, but with the maximum absolute value. The sign of $\theta$ indicates the direction of each link; that is, when the time lag is positive $\left(\theta >0\right)$, the direction of this link is from $j$ to $i$, and vice versa ~\cite{fan_network_2017}.

\subsubsection*{Null-model}
Next, we investigate the statistical significance of the link weights in the real networks by comparing to the shuffled surrogate network. In the surrogate network, to calculate link weight for each pair of nodes, we use two segment of data, each is corresponding to $30$ consecutive days starting from the first day of a month that is randomly selected from the period Jan. 1980-Dec. 2019, so that to destroy real correlations between two nodes in the temporal dimension.
Then we define the significant threshold $q$ as the $95\%$ highest value of the absolute weights for all links in the surrogate network. The link $(i,j)^m$ in the real network for a specific month $m$ is defined as significant if it is higher than $q$ or lower than $-q$, i.e., $|C_{i,j}^m|>q$. We find that the number of significant links for each month's network are dynamically changing with time as shown in Fig. S11.

\subsubsection*{Node degrees}
We define the degree for each global node as the number of significant links that connect to the Arctic nodes. We show heatmaps of node degrees for two specific months, i.e., the Feb. 2010 (Fig. \ref{PATH}a) and the Mar. 2019 (Fig. \ref{PATH}b). We observe higher node degrees in many regions, even in low latitudes, of the NH for Feb. 2010, comparing to that for Mar. 2019. We suppose it is related to the different phases of the AO.

\subsection*{Teleconnection path mining}
To identify the teleconnection path, we perform the \textit{shortest path} method of complex networks to find the optimal paths in our climate networks. A path is a sequence of nodes in which each node is adjacent to the next one, especially, in a directed network, the path can follow only the direction of an arrow. 
Here, our climate network is based on only one climate variable--air temperature at $850 hPa$ pressure level, and we select 726 nodes from the $10512$ nodes \cite{fan2022network,liu_teleconnections_2023}. 
For each climate network link $(i,j)^m$, we define its cost function value as
\begin{equation}
E_{i,j}^m = \frac{1}{|C_{i,j}^m|}.    
\end{equation}
The Dijkstra algorithm \cite{dijkstra_note_1959} was used to determine the directed optimal path between a source node $i$ and a sink node $j$ with the following constraints \cite{zhou_teleconnection_2015,liu_teleconnections_2023}: (i) the distance for every step is shorter than 1000km; (ii) link time delay $\theta \ge 0$; (iii) the sum cost function value for all collection of links through path $i \longrightarrow j$ is minimal. In this way, we identify the
optimal paths for information/energy/matter spreading in the two-dimensional
space.

\section*{Data availability}
The data represented in Figs. 2–5 are available as Source Data. All other data that support the plots within this paper and other findings of this study are available from the corresponding author upon reasonable request.

\section*{code availability}
The C++ and Python codes used for the analysis are available on GitHub: (\url{https://github.com/fanjingfang/DAWV}).\\

\section*{Acknowledgments}
The authors wish to thank T. Liu for his helpful suggestions. We acknowledge the support by the National Natural Science Foundation of China (Grant No. 12205025, 12275020, 12135003).

{\section*{Author Contributions}
J.M and J.F designed the research. J.M performed the analysis, J.M, J.F, U.S.B and J.K generated research ideas and discussed results, and contributed to writing the manuscript.}

\section*{Additional information}
Supplementary Information is available in the online version of the paper.

\section*{Competing interests}
The authors declare no competing  interests.
\bibliography{mylib_seaice}
\bibliographystyle{naturemag}
\end{document}